\documentclass[conference]{IEEEtran}
\IEEEoverridecommandlockouts
\usepackage{cite}
\usepackage{amsmath,amssymb,amsfonts}
\usepackage{algorithmic}
\usepackage{algorithm}
\usepackage{multirow}
\usepackage{graphicx}
\usepackage{textcomp}
\usepackage{xcolor}
\usepackage{hyperref}
\usepackage{cleveref}
\makeatletter
\let\MYcaption\@makecaption
\makeatother

\usepackage[font=footnotesize]{subcaption}

\makeatletter
\let\@makecaption\MYcaption
\makeatother

\def\BibTeX{{\rm B\kern-.05em{\sc i\kern-.025em b}\kern-.08em
    T\kern-.1667em\lower.7ex\hbox{E}\kern-.125emX}}

\newcommand{\etal}{\textit{et al}.}

\newcommand{\subject}{\mathcal{S}}
\newcommand{\dataset}{\mathcal{D}}
\newcommand{\loss}{\mathcal{L}}

\definecolor{darkpink}{rgb}{0.91, 0.33, 0.5}
\begin{document}

\title{Evaluating Fast Adaptability of Neural Networks for Brain-Computer Interface
}

\author{\IEEEauthorblockN{Anupam Sharma}
\IEEEauthorblockA{\textit{Department of Computer Science \& Engineering} \\
\textit{Indian Institute of Technology Gandhinagar}\\
Gujarat, India \\
sharmaanupam@iitgn.ac.in}
\and
\IEEEauthorblockN{Krishna Miyapuram}
\IEEEauthorblockA{\textit{Department of Cognitive \& Brain Sciences} \\
\textit{Indian Institute of Technology Gandhinagar}\\
Gujarat, India \\
kprasad@iitgn.ac.in}
}

\maketitle

\begin{abstract}
Electroencephalography (EEG)  classification is a versatile and portable technique for building non-invasive Brain-computer Interfaces (BCI). However, the classifiers that decode cognitive states from EEG brain data perform poorly when tested on newer domains, such as tasks or individuals absent during model training. Researchers have recently used complex strategies like Model-agnostic meta-learning (MAML) for domain adaptation. Nevertheless, there is a need for an evaluation strategy to evaluate the fast adaptability of the models, as this characteristic is essential for real-life BCI applications for quick calibration. We used motor movement and imaginary signals as input to Convolutional Neural Networks (CNN) based classifier for the experiments. Datasets with EEG signals typically have fewer examples and higher time resolution. Even though batch-normalization is preferred for Convolutional Neural Networks (CNN), we empirically show that layer-normalization can improve the adaptability of CNN-based EEG classifiers with not more than ten fine-tuning steps.
In summary, the present work (i) proposes a simple strategy to evaluate fast adaptability, and (ii) empirically demonstrate fast adaptability across individuals as well as across tasks with simple transfer learning as compared to MAML approach. 
\end{abstract}

\section{Introduction}
\label{sec:intro}

Brain-computer interface (BCI) systems provide an approach for devising efficient assistive technologies in rehabilitation and healthcare. Electroencephalography (EEG) being the most versatile technique to non-invasively record brain activity, makes it most suitable for building  BCI systems. The EEG technique records electrical signals in the brain by placing electrodes on the human scalp. Due to its feasibility in BCI systems, making efficient EEG classifiers is essential for wide-scale applications. In the present work, we focus on EEG classification with fast domain adaptation on signals for a well-known BCI application i.e. while performing body movements (motor movement) and imagining body movements (motor imagery).

Researchers have widely used traditional machine-learning approaches for EEG classification. However, these approaches need extensive feature extraction. With advancements in deep learning, deep neural networks (DNN) is being used widely due to its capability of extracting features implicitly~\cite{review, cnn_review}. In this domain, convolutional neural networks (CNN) have been the most widely used neural architecture for EEG classification~\cite{review, cnn_review}. 

The significant challenge in classifying EEG into corresponding cognitive states is the smaller dataset size, and variation in the signals across individuals. This variation leads to a degradation in the performance of EEG classifiers~\cite{review, variation} necessitating calibration or adaptation when applied to other domains like newer individuals or task activities. However, it is not feasible in real-time BCI applications to perform substantial calibration for each task and individual. Therefore, it is imperative to devise techniques for fast adaptability of EEG classifiers.

Inspired by computer vision, most CNN-based EEG classifiers use batch-norm. However, EEG datasets are smaller, necessitating a tradeoff between batch size and the number of iterations per epoch. In addition, for fast calibration on new individuals or tasks, collecting large samples is not feasible in real-time applications, making batch-norm inappropriate for fast adaptability. However, EEG signals are long sequences in the time domain, which is enough to calculate statistics for normalization, making layer-norm perfectly applicable for such applications. We empirically show that layer-norm helps adapt EEG classifiers notably faster than the classifiers with batch-norm on new individuals as well as on new activities.

In summary, it is important that BCI systems take minimal time for adaptation in practical applications when used for newer individuals or activities with few samples. Therefore, in this work:

\begin{enumerate}
    \item We show that changes in normalization techniques following the characteristics of EEG signals can help improve adaptability with vanilla transfer learning.
    \item We propose an effective evaluation strategy where we evaluate the fast adaptability of two popular training strategies, MAML and transfer learning, by limiting the fine-tuning iterations to ten, and we show that transfer learning adapts much faster than MAML. Moreover, we evaluate fast adaptability not only on newer individuals but also on signals on newer task / activities.
\end{enumerate}

\section{Related Work}
\label{sec:related_work}

To tackle the problem of variations in EEG signals, several approaches have been proposed. In~\cite{gcram}, authors proposed a Graph-based Convolutional Recurrent Attention Model (G-CRAM) model where the authors used a graph structure to represent the positioning of EEG electrodes and then used a convolutional recurrent attention model to learn EEG features. In~\cite{fusion}, authors used the concept of weighted feature fusion in CNN for motor imagery EEG decoding. Researchers have also used feature extraction methods before using deep neural networks. In~\cite{csp_cnn}, authors used a combination of common spatial patterns and CNN to develop the decoding model. In addition to generalizing EEG decoders across individuals, researchers have also explored transfer-learning techniques to improve the decoding performance~\cite{review, anotherTL}. In~\cite{eegsym}, the authors have introduced the inception module and residual connection in their model architecture and pre-trained the model on large data. The authors curated the dataset of 280 individuals by collecting the data from common electrodes from multiple datasets. Researchers have also explored meta-learning methods, which observe learning approaches on different tasks and then use this experience to solve a new task faster~\cite{auto_ml_meta}. One such meta-learning algorithm is model-agnostic meta-learning (MAML)~\cite{maml} which researchers have started exploring recently~\cite{mamlbci, mamlbci1, mamlbci2, mamlbci3, mamlbci4}. For example, in~\cite{mamlbci}, authors assumed different individuals as a different classification tasks and used MAML for adaptation to newer individuals.
\section{Methodology}
\label{sec:methodology}

The principle idea of our work is to investigate the fast adaptability of EEG decoding deep neural networks when applied to the signals of newer individuals or signals recorded while performing different activities. Hence, we can have two adaptability tasks:

\begin{itemize}
    \item \textbf{Across individual:} Introducing signals of newer individuals, making it the same classification task but on signals of different individuals (In the subsequent section, we often use the term ``subject" to refer to an individual).
    \item  \textbf{Across Task activity:} Introducing signals recorded on newer activities, changing the set of labels to classify, making it a different classification task altogether.
\end{itemize}

%-------------------------------------------------------------------------
\subsection{Training strategies}

To evaluate the fast adaptability, we compared two training strategies by limiting the number of iterations in fine-tuning to $10$ and compared the classification performance. In this section, we elaborate on our two training strategies used for domain adaptation experiments for EEG classification.

\textbf{Model-agnostic meta-learning (MAML)}~\cite{maml}. This strategy is specifically applicable in the few-shot meta-learning setup and is well adapted in the BCI domain~\cite{mamlbci,mamlbci1,mamlbci2}. Its goal is to learn parameters so the model can adapt to new tasks using only a few data points and iterations. In this work, we have adapted MAML for EEG classification across subjects as in~\cite{mamlbci}. Considering the model with parameters $\theta$, the algorithm first samples a batch of $N$ subjects $\subject_i$. For each subject $\subject_i$, the algorithm samples support set $\dataset = \{\mathbf{x}^j, \mathbf{y}^j\}$ and a query set $\dataset' = \{\mathbf{x}^j, \mathbf{y}^j\}$ representing an $n$-way $k$-shot classification task and creates a copy of subject-specific parameters $\theta_i$. These parameters are learned for each subject $\subject_i$ separately on the support set for one or more gradient descent steps. For example, for a subject $\subject_i$ and one step of gradient descent, the parameters would be updated as:

\begin{equation}
    \label{eq:inner_sgd}
    \phi_{i}^{1} = \theta - \alpha \nabla_\theta \loss_{\subject_i,\dataset} (\theta)
\end{equation}

Finally, the model parameters are updated by evaluating gradients of loss on query-set $\dataset'$ using updated parameters $\phi_{i}^{n}$ across subjects. For example, using gradient descent, the model parameters are updated as:

\begin{equation}
    \label{eq:meta_update}
    \begin{split}
        \theta &= \theta - \beta \frac{1}{N} \sum_{i=1}^{N} \nabla_\theta \loss_{\subject_i,\dataset'}(\phi_{i}^{n})\\
    &= \theta - \beta \frac{1}{N} \sum_{i=1}^{N} \nabla_{\phi_{i}^{n}} \loss_{\subject_i,\dataset'}(\phi_{i}^{n}) \nabla_\theta \phi_{i}^{n}
    \end{split}
\end{equation}

\Cref{eq:meta_update} is referred as meta-update. However, the term $\nabla_\theta \phi_{i}^{n}$ involves higher-order derivatives:

\begin{equation}
    \label{eq:hesian}
    \begin{split}
        \nabla_\theta \phi_{i}^{n} &= \prod_{m=1}^{n} \nabla_{\phi_{i}^{m-1}} \phi_{i}^{m}\\
        &= \prod_{m=1}^{n} (\mathbf{\mathit{I}} - \alpha \nabla_{\phi_{i}^{m-1}}^2 \loss_{\subject_i,\dataset} (\phi_{i}^{m-1}))
    \end{split}
\end{equation}
The terms involving higher-order derivatives are costly to compute. Therefore, we adapt the first-order approximation of MAML where the algorithm ignores the terms involving higher-order derivatives and \cref{eq:meta_update} reduces to:

\begin{equation}
    \label{eq:fomaml_meta_update}
    \theta = \theta - \beta \frac{1}{N} \sum_{i=1}^{N} \nabla_{\phi_{i}^{n}} \loss_{\subject_i,\dataset'}(\phi_{i}^{n})
\end{equation}

However, the approximate algorithm is still computationally expensive as it involves nested loops. The model is pre-trained by repeating this process for multiple iterations. The process is outlined in \cref{alg: maml}.

\begin{algorithm}
    \caption{MAML adapted for EEG classification~\cite{maml, mamlbci}}
    \label{alg: maml}
    \begin{algorithmic}[1]
        \REQUIRE $p(\subject)$: Distributions over subjects
        \REQUIRE $\alpha, \beta$: inner-loop and meta learning rate respectively
        \STATE Randomly initialize model parameters, $\theta$
        \WHILE{not done}
            \STATE Sample batch of subjects $\subject_i \sim p(\subject)$  of size $N$
            \FORALL{$\subject_i$}
                \STATE Sample support data points $\dataset = \{\mathbf{x}^j, \mathbf{y}^j\}$ from $\subject_i$ 
                \STATE Compute adapted parameters, $\phi_{i}^{n}$, using $n$ gradient descent steps and loss function $\loss_{\subject_i,\dataset}(\theta)$
                \STATE Sample query data points $\dataset' = \{\mathbf{x}^j, \mathbf{y}^j\}$ from $\subject_i$
                \STATE Compute loss $\loss_{\subject_i,\dataset'}(\phi_{i}^{n})$ for meta-update
            \ENDFOR
            \STATE Perform meta-update using \cref{eq:fomaml_meta_update}
        \ENDWHILE
    \end{algorithmic}
\end{algorithm}

 \textbf{Transfer learning}. Research communities from different domains have used this strategy for various purposes. Considering the example of the classification task, in transfer learning, we first train the model to perform $n$-way classification on a large dataset, including samples from multiple subjects. Given a new subject representing a new $m$-way classification task, we fine-tune the pre-trained model on a new task for a few iterations by updating all the parameters on samples of the new subject. 

%-------------------------------------------------------------------------
\subsection{Model architecture}

We used a variant of EEGNet~\cite{eegnet} architecture for our experiments. EEGNet is a compact convolutional neural network (CNN) for decoding EEG signals. We selected CNN as it has been widely used for EEG classification~\cite{review, cnn_review}. Nevertheless, recent works have also used transformers-based models~\cite{transformer_1, transformer_2, transformer_3}, but it is not suitable when we have limited data, as in our case.

Inspired by computer vision, most CNN-based EEG classifiers use batch-norm. However, EEG datasets are smaller, necessitating a tradeoff between batch size and the number of iterations per epoch. Batch-norm is not preferable for low batch size. However, EEG signals are sequences with long sizes, making it a perfect application of layer-norm. Moreover, layer-norm is independent of batch size.
Furthermore, as our objective is to adapt to newer tasks quickly, we should refrain from using the statistics calculated during the pre-training phase. We have used an architecture similar to EEGNet; instead, we replaced batch-norm with layer-norm.  We empirically show that a model with layer-norm works better than a model with batch-norm in our experiments. The model architecture is described in \Cref{tab:arch} and remained fixed in all the experiments.

\begin{table}
\caption{Model Architecture}
\centering
\resizebox{7cm}{!}{%
\begin{tabular}{lllll}
\hline
Block              & \multicolumn{1}{l}{Layer} & \multicolumn{1}{l}{\#filters} & \multicolumn{1}{l}{Size} & \multicolumn{1}{l}{Others} \\ \hline
\multirow{7}{*}{1} & Conv2D          & 8  & 1,64 & padding=same  \\
                   & LayerNorm       &    &      &               \\
                   & DepthWiseConv2D & 16 & 64,1 & padding=valid \\
                   & LayerNorm       &    &      &               \\
                   & Elu Activation  &    &      &               \\
                   & AveragePool2D   &    & 1,4  & strides=1,4   \\
                   & DropOut         &    &      & p=0.25         \\ \cline{1-5}
\multirow{6}{*}{2} & SeparableConv2D            & 16                             & 1,16                      & padding=same                \\
                   & LayerNorm       &    &      &               \\
                   & Elu Activation  &    &      &               \\
                   & AveragePool2D   &    & 1,8  &               \\
                   & DropOut         &    &      & p=0.25         \\
                   & Flatten         &    &      &               \\ \cline{1-5}
3                  & Dense           &    & 160  &               \\ \cline{1-5}
\end{tabular}%
}

\label{tab:arch}
\end{table}
\section{Experiments}
\label{sec:experiments}

In this section, we discuss the effect of components of the training pipeline, like normalization layers and training approach on the adaptability of EEG classifiers. Our implementation is based on PyTorch~\cite{pytorch} for Python. The MAML implementation is highly inspired by the work of Long~\cite{MAML_Pytorch}. The code for experiments is available online \footnote{The code is available at \href{https://github.com/anp-scp/fast_bci/}{github.com/anp-scp/fast\_bci/}}.

%-------------------------------------------------------------------------
\subsection{Dataset and pre-processing}
\label{subsec: dataset_and_pre_processing}

We used Physionet's EEG Motor Movement/Imagery Dataset~\cite{physionet} for our experiments. The dataset includes EEG signals of $109$ individuals recorded while performing the following body activities (specified as Task in the dataset):

\begin{itemize}
    \item \textbf{Activity 1 (A1):} Open and close left or right fist
    \item \textbf{Activity 2 (A2):} Imagine opening and closing left or right fist
    \item \textbf{Activity 3 (A3):} Open and close both fists or both feet
    \item \textbf{Activity 4 (A4):} Imagine opening and closing both fists or both feet
\end{itemize}

The signals were recorded using an EEG device with $64$ electrodes returning $64$ time-domain EEG signals with a sampling frequency of $160$ Hz. For pre-precessing, we used MNE-python~\cite{mne}, a library for Python language for EEG and MEG analysis. Moreover, the dataset is readily available within the MNE-python library and can be directly used via the interface available in the library. We performed the following pre-processing steps:

\begin{enumerate}
    \item We applied a ``firwin" band-stop filter allowing frequencies outside $7-30$ Hz following~\cite{gamma}.
    \item We segmented all the time domain signals such that each segment starts $1$ second before and ends $4$ seconds after stimulus onset.
    \item Use the first $2$ seconds of the segments, returning the matrix of shape $64\times 321$. Here, 6$4$ represents each channel, and $321$ represents time points (samples with time points less than $321$ were dropped).
\end{enumerate}

For all the experiments, we used the data of subjects $1-87$ as the training set and $88-98$ as the validation set while pre-training. We evaluated fast adaptability on the data of subjects $99-109$.

%-------------------------------------------------------------------------
\subsection{Batch-norm vs layer-norm}
\label{subsec:batch_layer}

In this experiment, we investigated the fast adaptability of a model trained using the transfer-learning strategy with batch-norm vs layer-norm. For each model architecture (i.e., with batch norm and layer norm), we performed binary classification between feet and fist for activity 4 as described in the \cref{subsec: dataset_and_pre_processing}. For the hyperparameter search, we tried learning rate $[0.01, 0.001]$ and batch size $[16, 32, 64]$, and we selected the one with the best validation accuracy. For both architectures, the learning rate was selected as $0.001$ and the batch size as $16$. All the other hyperparameters remain fixed, as described in the \Cref{tab:best-hp-transfer}.

%-------------------------------------------------------------------------
\subsection{Adaptability across individuals}
\label{subsec:across_individual}

In this experiment, we investigate the adaptability of models on newer individuals. We compare two training strategies: MAML and transfer learning. For MAML, we considered data for each individual as a different task as in~\cite{mamlbci} and performed binary classification for each activity separately. For the hyperparameter search, we tried 5 and 10 adapt steps for the inner loop in MAML and 0.01 and 0.001 for learning rates, out of which we selected the one with the best validation accuracy. The best hyperparameters are reported in \Cref{tab:best-hp-maml} and \Cref{tab:best-hp-transfer}. We then fine-tune the pre-trained models on the data of subjects $99-109$ separately to evaluate the fast adaptability of the models.

\begin{table}
\centering
\caption{Best pre-training hyperparameters involving MAML for experiments evaluating adaptability across individuals and activities. The values inside brackets ({\color{darkpink}typeset in pink}) denote the activity. For example, $0.001$ ({\color{darkpink}A4}) denotes that the value $0.001$ is for Activity $4$}
\label{tab:best-hp-maml}
\resizebox{\columnwidth}{!}{%
\begin{tabular}{ll}
\hline
\textbf{Parameter} & \textbf{Value} \\ \hline
Dropout & 0.25 \\
Number of samples per class in support set & 10 \\
Number of samples per class in query set & 11 \\
Number of subjects in a batch per epoch & 4 \\
Loss function & Cross entropy loss \\
Optimizer in inner loop & Gradient Descent \\
Meta-optimizer & Adam \\
Inner loop learning rate & 0.01 ({\color{darkpink}A1, A2, A3}), 0.001 ({\color{darkpink}A4}) \\
Meta-learning rate & 0.001 ({\color{darkpink}A1}), 0.01 ({\color{darkpink}A2, A3, A4}) \\
Inner loop adapt steps & 10 ({\color{darkpink}A1, A2}), 5 ({\color{darkpink}A3, A4}) \\ \hline
\end{tabular}%
}
\end{table}

\begin{table}
\centering
\caption{Best pre-training hyperparameters involving transfer learning for experiments evaluating adaptability across individuals and activities. The values inside brackets ({\color{darkpink}typeset in pink}) denote the activity. For example, $0.001$ ({\color{darkpink}A4}) denotes that the value $0.001$ is for Activity $4$}
\label{tab:best-hp-transfer}
\begin{tabular}{ll}
\hline
\textbf{Parameter} & \textbf{Value} \\ \hline
Dropout & 0.25 \\
Number of samples per class & 21 \\
Loss function & Cross entropy loss \\
Optimizer & Adam \\
Learning rate & 0.001 ({\color{darkpink}A1, A2, A3, A4}) \\
Batch size & 32 ({\color{darkpink}A1, A3}), 16 ({\color{darkpink}A2, A4})  \\ \hline
\end{tabular}%

\end{table}

%-------------------------------------------------------------------------
\subsection{Adaptability across task activities}

In this experiment, we investigate the adaptability of EEG decoding models on signals of newer individuals recorded on newer activities. For example, a model trained on data from Activity $1$ is adapted to the data from Activity $4$. For this experiment, we use the models pre-trained in the experiment described in \Cref{subsec:across_individual} and fine-tuned them on the data from another activity.

%-------------------------------------------------------------------------
\subsection{Evaluation metrics}
We evaluated the adaptability of the models by fine-tuning the pre-trained model on the data of subjects $99-109$ separately. We don't tune hyper-parameters while fine-tuning.

In the case of experiments involving vanilla transfer learning, we considered $10$ samples per class during fine-tuning and $11$ per class to test the model for each subject. All the model parameters were fine-tuned with a learning rate of $0.001$ for $10$ iterations optimized using the Adam optimizer. We used a similar paradigm for evaluation in MAML experiments, except that Gradient Descent was used to fine-tune the model parameters. In the case of MAML, the learning rate was the same as the inner loop learning rate during pre-training for experiments evaluating adaptability across individuals and $0.001$ for experiments evaluating adaptability across task activities. For this, we iterated the inner loop without iterating over the outer loop in \cref{alg: maml}, avoiding any meta-update. The number of samples in the support set was set to $10$ per class, and the number of samples in the query set to $11$. 

Finally, we reported the accuracy on the test samples (query set for MAML) averaged across subjects $99-109$. As the amount of data we had was small, we performed the fine-tuning for $100$ runs and reported the mean and standard deviation of the accuracy.

%----------------------------------------------
%Figures

\begin{figure*}
  \centering
  \begin{subfigure}{0.4\linewidth}
    \includegraphics[width=\linewidth]{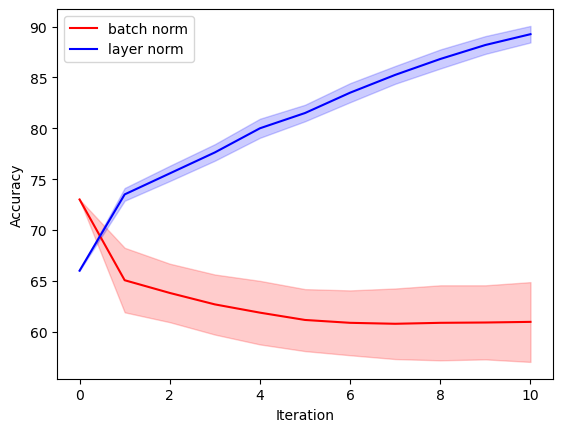}
    \caption{Trend of training accuracy during fine-tuning}
    \label{fig:batch_layer-a}
  \end{subfigure}
  \begin{subfigure}{0.4\linewidth}
    \includegraphics[width=\linewidth]{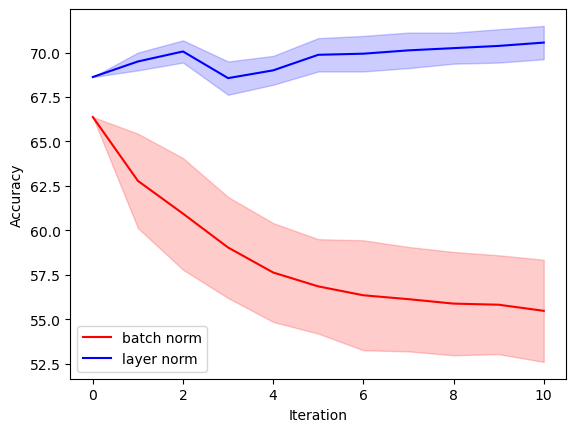}
    \caption{Trend of testing accuracy during fine-tuning}
    \label{fig:batch_layer-b}
  \end{subfigure}
  \caption{Performance analysis (in accuracy) of models with batch-norm and with layer-norm}
  \label{fig:batch_layer}
\end{figure*}

\begin{figure*}
  \centering
  \begin{subfigure}{0.4\linewidth}
    \includegraphics[width=\linewidth]{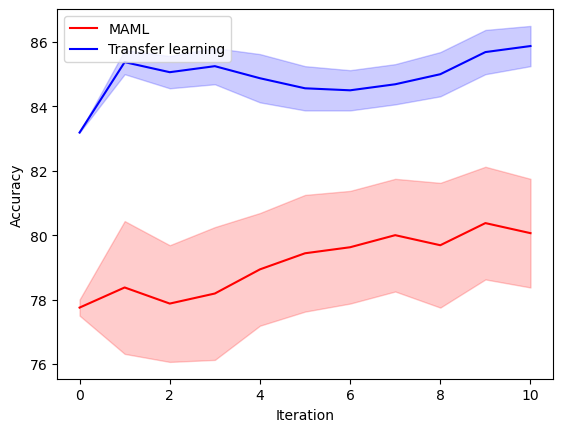}
    \caption{Trend of test accuracy during fine-tuning on activity $1$}
    \label{fig:across_individual-a}
  \end{subfigure}
  \begin{subfigure}{0.4\linewidth}
    \includegraphics[width=\linewidth]{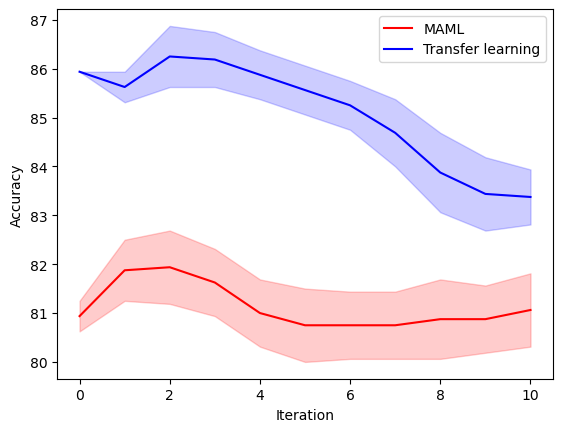}
    \caption{Trend of test accuracy during fine-tuning on activity $2$}
    \label{fig:across_individual-b}
  \end{subfigure}
  \begin{subfigure}{0.4\linewidth}
    \includegraphics[width=\linewidth]{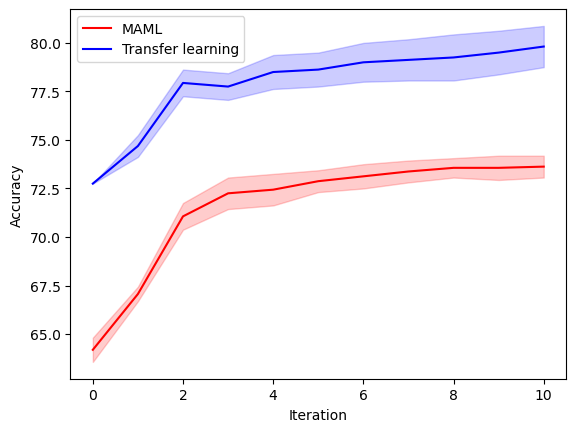}
    \caption{Trend of test accuracy during fine-tuning on activity $3$}
    \label{fig:across_individual-c}
  \end{subfigure}
  \begin{subfigure}{0.4\linewidth}
    \includegraphics[width=\linewidth]{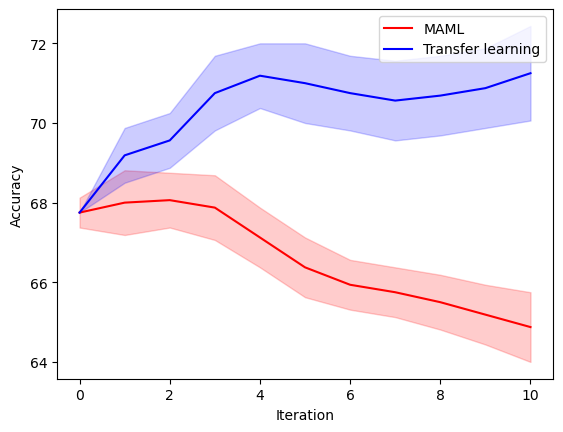}
    \caption{Trend of test accuracy during fine-tuning on activity $4$}
    \label{fig:across_individual-d}
  \end{subfigure}
  \caption{Performance analysis (in accuracy) of training strategies when adapted to newer individuals}
  \label{fig:across_individual}
\end{figure*}

\begin{figure*}
  \centering
  \begin{subfigure}{0.25\linewidth}
    \includegraphics[width=\linewidth]{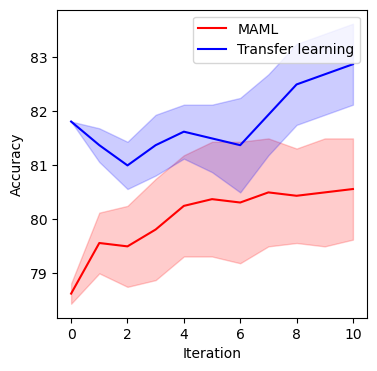}
    \caption{Adapting from Activity $1$ to $2$}
    \label{fig:t1_to_all-a}
  \end{subfigure}
  \begin{subfigure}{0.25\linewidth}
    \includegraphics[width=\linewidth]{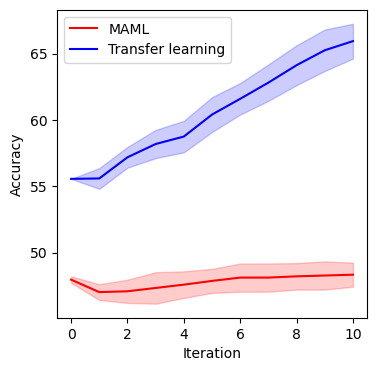}
    \caption{Adapting from Activity $1$ to $3$}
    \label{fig:t1_to_all-b}
  \end{subfigure}
  \begin{subfigure}{0.25\linewidth}
    \includegraphics[width=\linewidth]{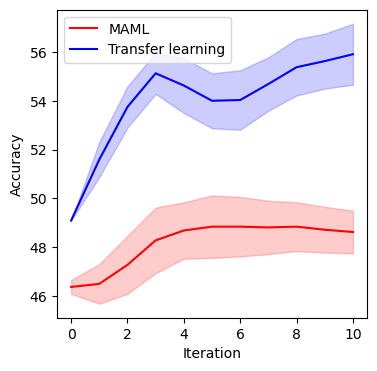}
    \caption{Adapting from Activity $1$ to $4$}
    \label{fig:t1_to_all-c}
  \end{subfigure}
  %%---------------------------
  \begin{subfigure}{0.25\linewidth}
    % \fbox{\includegraphics[width=\linewidth]{batch_vs_layer_train.png}}
    \includegraphics[width=\linewidth]{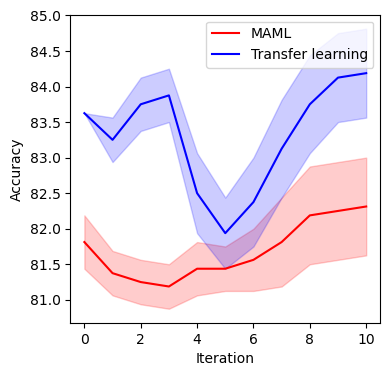}
    \caption{Adapting from Activity $2$ to $1$}
    \label{fig:t1_to_all-d}
  \end{subfigure}
  \begin{subfigure}{0.25\linewidth}
    \includegraphics[width=\linewidth]{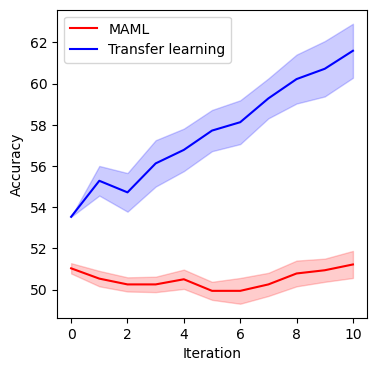}
    \caption{Adapting from Activity $2$ to $3$}
    \label{fig:t1_to_all-e}
  \end{subfigure}
  \begin{subfigure}{0.25\linewidth}
    \includegraphics[width=\linewidth]{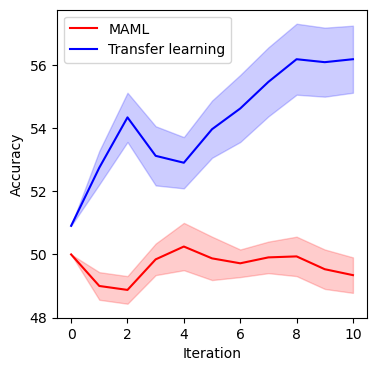}
    \caption{Adapting from Activity $2$ to $4$}
    \label{fig:t1_to_all-f}
  \end{subfigure}
  %------------------------------------
  \begin{subfigure}{0.25\linewidth}
    \includegraphics[width=\linewidth]{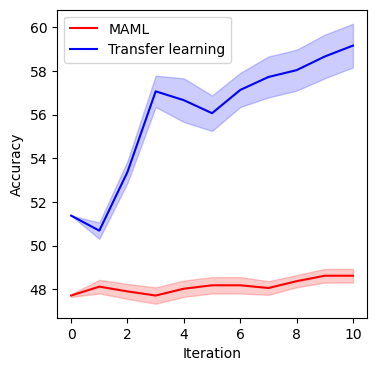}
    \caption{Adapting from Activity $3$ to $1$}
    \label{fig:t1_to_all-g}
  \end{subfigure}
  % \hfill
  \begin{subfigure}{0.25\linewidth}
    \includegraphics[width=\linewidth]{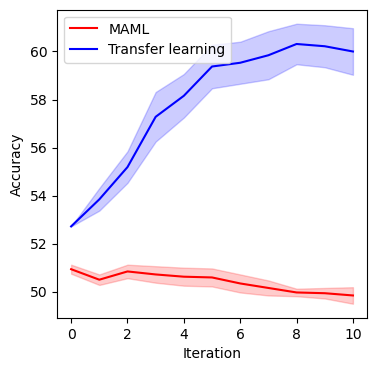}
    \caption{Adapting from Activity $3$ to $2$}
    \label{fig:t1_to_all-h}
  \end{subfigure}
  \begin{subfigure}{0.25\linewidth}
    \includegraphics[width=\linewidth]{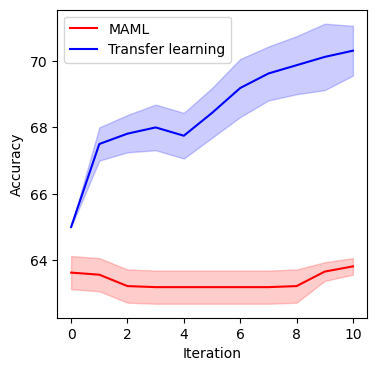}
    \caption{Adapting from Activity $3$ to $4$}
    \label{fig:t1_to_all-i}
  \end{subfigure}
  %%--------------------------------
  \begin{subfigure}{0.25\linewidth}
    \includegraphics[width=\linewidth]{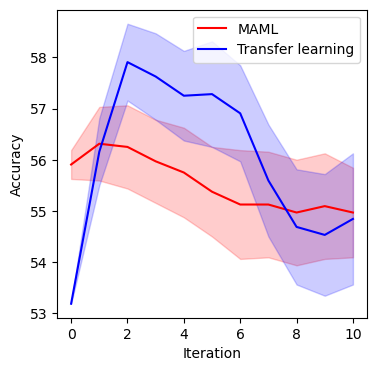}
    \caption{Adapting from Activity $4$ to $1$}
    \label{fig:t1_to_all-j}
  \end{subfigure}
  % \hfill
  \begin{subfigure}{0.25\linewidth}
    \includegraphics[width=\linewidth]{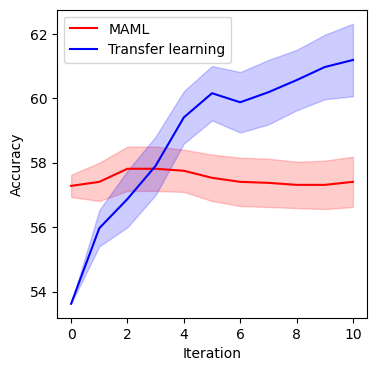}
    \caption{Adapting from Activity $4$ to $2$}
    \label{fig:t1_to_all-k}
  \end{subfigure}
  \begin{subfigure}{0.25\linewidth}
    \includegraphics[width=\linewidth]{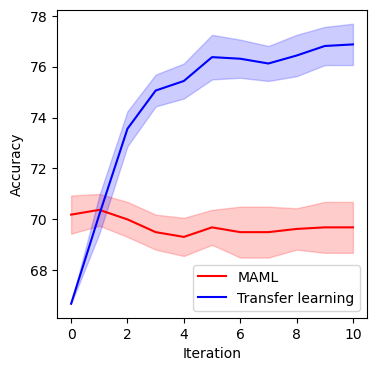}
    \caption{Adapting from Activity $4$ to $3$}
    \label{fig:t1_to_all-l}
  \end{subfigure}
  \caption{Performance analysis (test accuracy) of training strategies when model trained on one activity is adapted to other activities}
  \label{fig:t1_to_all}
\end{figure*}

\section{Results and Discussion}
\label{sec:discussion}

In this paper, we evaluated the fast adaptability of motor movement and imagery EEG decoding models from the normalization technique and training strategy perspective. In the upcoming subsection, we discuss the results of all the experiments we performed.

\subsection{Batch-norm vs layer-norm}

We compared the performance of models with batch-norm and layer-norm when adapted to the EEG signals of newer individuals. Though batch-norm is widely used with CNN models, we can observe in \Cref{fig:batch_layer} that the model with batch-norm layer cannot adapt to the signals of newer individuals, whereas the model with layer-norm layer adapts well within 10 iterations. Moreover, the training and testing accuracy degrades as we continue fine-tuning the model with the batch-norm. Such decline can be attributed to the problem of variations in EEG signals across each individual or activity~\cite{review, variation}. Hence, it is essential that we calculate data statistics for normalization from the new data and not from the training data. Even though the statistics change during the fine-tuning phase, it is not favorable for fast adaptability, especially when we have fewer samples. For practical applications, we cannot have large samples for quick calibration, and these results show that layer-norm is a better choice for normalization than batch-norm for EEG classification in such scenarios.

\subsection{Adaptability across individuals}

We further compared the performance of models trained with two popular knowledge transfer strategies, MAML and transfer learning, when adapted to signals of newer individuals. Even though MAML was developed for adaptation, our results show that is not necessarily the case for EEG decoding. In \Cref{fig:across_individual}, we can observe that as we fine-tune the models on newer individuals, the model trained with transfer learning adapts considerably faster than the model trained with MAML for all the activities mentioned in \Cref{subsec: dataset_and_pre_processing}. Notably, for activity $2$, the model trained with transfer learning attains its maximum at iteration $2$. Also, the model's initial parameters (at iteration $0$ during fine-tuning) perform better when trained with transfer learning than MAML. The results here are opposite to the one reported by Li \etal~\cite{mamlbci}, where MAML was found to perform better. Moreover, our approach attains similar performance without excluding a significant dataset section, as in~\cite{mamlbci}. We also found that when models are adapted to newer individuals for the same set of labels, the model trained with transfer learning performs reasonably well within $5$ iterations, as evident in \Cref{fig:across_individual}.

Finally, in \Cref{tab:result}, we compare the performance of our approach and the ones reported by other recent state-of-the-art approaches. Few cells in the table are left blank as the authors did not perform corresponding experiments. None except us performed the experiments on Activity $1$ and Activity $3$. For activity $2$, our approach performed the best when tested without adaptation. Moreover, after adaptation, our approach performed slightly lower than EEGSym~\cite{eegsym}, even though the amount of pre-training data for EEGSym~\cite{eegsym} is notably higher than our pre-training data. For activity $4$, our approach performed very close to Li \etal's~\cite{mamlbci} approach without filtering any samples. Li \etal~\cite{mamlbci} reached an accuracy of $79.7$ after filtering nearly $80\%$ of samples. In summary, our approach with a simple model and training strategy highly competes with the existing state-of-the-art models on Physionet's EEG Motor Movement/Imagery Dataset~\cite{physionet} specifically in a low data environment.

\begin{table}
\caption{Comparison of our performance (test accuracy) and the performance reported by other recent state-of-the-art approaches on Physionet's EEG Motor Movement/Imagery Dataset~\cite{physionet}. Certain cells are blank, as the corresponding authors did not reported them.}
\centering
\resizebox{\columnwidth}{!}{%
\begin{tabular}{llll}
\hline
Activity & \multicolumn{1}{c}{Model} & \multicolumn{1}{c}{Before adaptation} & \multicolumn{1}{c}{After adaptation} \\ \hline
1 & \textbf{Ours} & 83.18 $\pm$ 0 & 85.88 $\pm$ 0.6 \\ \hline
\multirow{6}{*}{2} & DG-CRAM~\cite{gcram} & 74.71 $\pm$ 4.19 & - \\
 & EEGSym~\cite{eegsym} & - & 88.6 $\pm$ 9.0 \\
 & Li \etal~\cite{mamlbci} & - & 80.6 \\
 & s-CTrans~\cite{transformer_1} & 83.31 & - \\
 & \textbf{Ours} & 85.91 $\pm$ 0 & 86.28 $\pm$ 0.63 \\ \hline
3 & \textbf{Ours} & 72.73 $\pm$ 0 & 79.81 $\pm$ 1.07 \\ \hline
\multirow{3}{*}{4} & Li \etal~\cite{mamlbci}& - & 79.7 \\
 & \textbf{Ours} & 67.73 $\pm$ 0 & 71.22 $\pm$ 1.2 \\ \cline{1-4} 
\end{tabular}%
}

\label{tab:result}
\end{table}

\subsection{Adaptability across task activities}

\Cref{fig:t1_to_all} compares the models' performance with MAML and transfer learning when the model trained on signals of activity $\mathit{i}$ is fine-tuned to activity $\mathit{j}$ (where $\mathit{i} \neq \mathit{j}$). We can observe that in all the cases, the model trained with transfer learning performs better. However, the performance is low when the activities are not similar (The activity set {1,2} is different from the activity set {3,4} as they involve different body movements). Nevertheless, it is interesting that in such cases, models trained with MAML either cannot cross the $50\%$ chance level or remain very near the chance level, as seen in \cref{fig:t1_to_all-b}, \Cref{fig:t1_to_all-c}, \Cref{fig:t1_to_all-e}, \Cref{fig:t1_to_all-f}, \Cref{fig:t1_to_all-g} and \Cref{fig:t1_to_all-h}. It is only in the case when the models trained on activity $4$ are adapted to other activities that the initial parameters of the model with MAML provide better performance far from the chance level (\Cref{fig:t1_to_all-j} and \Cref{fig:t1_to_all-k}). However, as we fine-tune the models for 10 steps, transfer learning adapts quickly and beats MAML. The better adaptability of EEG decoding models trained with transfer learning, as seen in our experiments, shows that transfer learning remains the better option for cross-individual, cross-activity motor movement and imagery EEG decoding.
\section{Conclusion}
\label{sec:conclusion}

This work provides insights on evaluating models and training strategies for fast adaptability for across-individual and across-activity motor movement and imagery EEG decoding. We also suggest an architectural change in the models for better adaptability. Our proposed architecture i.e. CNN with layer-norm and transfer learning shows faster adaptation than recently proposed MAML approach, and thereby will be suitable for fast calibration of real-time BCI systems.

\bibliographystyle{IEEEtran}
\bibliography{IEEEabrv,mybibfile}

\begin{thebibliography}{10}
\providecommand{\url}[1]{#1}
\csname url@samestyle\endcsname
\providecommand{\newblock}{\relax}
\providecommand{\bibinfo}[2]{#2}
\providecommand{\BIBentrySTDinterwordspacing}{\spaceskip=0pt\relax}
\providecommand{\BIBentryALTinterwordstretchfactor}{4}
\providecommand{\BIBentryALTinterwordspacing}{\spaceskip=\fontdimen2\font plus
\BIBentryALTinterwordstretchfactor\fontdimen3\font minus \fontdimen4\font\relax}
\providecommand{\BIBforeignlanguage}[2]{{%
\expandafter\ifx\csname l@#1\endcsname\relax
\typeout{** WARNING: IEEEtran.bst: No hyphenation pattern has been}%
\typeout{** loaded for the language `#1'. Using the pattern for}%
\typeout{** the default language instead.}%
\else
\language=\csname l@#1\endcsname
\fi
#2}}
\providecommand{\BIBdecl}{\relax}
\BIBdecl

\bibitem{review}
\BIBentryALTinterwordspacing
H.~Altaheri, G.~Muhammad, M.~Alsulaiman, S.~U. Amin, G.~A. Altuwaijri, W.~Abdul, M.~A. Bencherif, and M.~Faisal, ``Deep learning techniques for classification of electroencephalogram (eeg) motor imagery (mi) signals: a review,'' \emph{Neural Computing and Applications}, vol.~35, no.~20, pp. 14\,681--14\,722, Jul 2023. [Online]. Available: \url{https://doi.org/10.1007/s00521-021-06352-5}
\BIBentrySTDinterwordspacing

\bibitem{cnn_review}
\BIBentryALTinterwordspacing
A.~Craik, Y.~He, and J.~L. Contreras-Vidal, ``Deep learning for electroencephalogram (eeg) classification tasks: a review,'' \emph{Journal of Neural Engineering}, vol.~16, no.~3, p. 031001, apr 2019. [Online]. Available: \url{https://dx.doi.org/10.1088/1741-2552/ab0ab5}
\BIBentrySTDinterwordspacing

\bibitem{variation}
V.~Jayaram, M.~Alamgir, Y.~Altun, B.~Sch{\"o}lkopf, and M.~Grosse-Wentrup, ``Transfer learning in brain-computer interfaces,'' \emph{IEEE Computational Intelligence Magazine}, vol.~11, no.~1, pp. 20--31, 2016.

\bibitem{gcram}
D.~Zhang, K.~Chen, D.~Jian, and L.~Yao, ``Motor imagery classification via temporal attention cues of graph embedded eeg signals,'' \emph{IEEE Journal of Biomedical and Health Informatics}, vol.~24, no.~9, pp. 2570--2579, 2020.

\bibitem{fusion}
S.~U. Amin, M.~Alsulaiman, G.~Muhammad, M.~A. Bencherif, and M.~S. Hossain, ``Multilevel weighted feature fusion using convolutional neural networks for eeg motor imagery classification,'' \emph{IEEE Access}, vol.~7, pp. 18\,940--18\,950, 2019.

\bibitem{csp_cnn}
\BIBentryALTinterwordspacing
X.~Zhu, P.~Li, C.~Li, D.~Yao, R.~Zhang, and P.~Xu, ``Separated channel convolutional neural network to realize the training free motor imagery bci systems,'' \emph{Biomedical Signal Processing and Control}, vol.~49, pp. 396--403, 2019. [Online]. Available: \url{https://www.sciencedirect.com/science/article/pii/S1746809418303264}
\BIBentrySTDinterwordspacing

\bibitem{anotherTL}
\BIBentryALTinterwordspacing
K.~Zhang, N.~Robinson, S.-W. Lee, and C.~Guan, ``Adaptive transfer learning for eeg motor imagery classification with deep convolutional neural network,'' \emph{Neural Networks}, vol. 136, pp. 1--10, 2021. [Online]. Available: \url{https://www.sciencedirect.com/science/article/pii/S0893608020304305}
\BIBentrySTDinterwordspacing

\bibitem{eegsym}
S.~Pérez-Velasco, E.~Santamaría-Vázquez, V.~Martínez-Cagigal, D.~Marcos-Martínez, and R.~Hornero, ``Eegsym: Overcoming inter-subject variability in motor imagery based bcis with deep learning,'' \emph{IEEE Transactions on Neural Systems and Rehabilitation Engineering}, vol.~30, pp. 1766--1775, 2022.

\bibitem{auto_ml_meta}
\BIBentryALTinterwordspacing
J.~Vanschoren, \emph{Meta-Learning}.\hskip 1em plus 0.5em minus 0.4em\relax Cham: Springer International Publishing, 2019, pp. 35--61. [Online]. Available: \url{https://doi.org/10.1007/978-3-030-05318-5\_2}
\BIBentrySTDinterwordspacing

\bibitem{maml}
C.~Finn, P.~Abbeel, and S.~Levine, ``Model-agnostic meta-learning for fast adaptation of deep networks,'' in \emph{Proceedings of the 34th International Conference on Machine Learning - Volume 70}, ser. ICML'17.\hskip 1em plus 0.5em minus 0.4em\relax JMLR.org, 2017, p. 1126–1135.

\bibitem{mamlbci}
D.~Li, P.~Ortega, X.~Wei, and A.~Faisal, ``Model-agnostic meta-learning for eeg motor imagery decoding in brain-computer-interfacing,'' in \emph{2021 10th International IEEE/EMBS Conference on Neural Engineering (NER)}, 2021, pp. 527--530.

\bibitem{mamlbci1}
\BIBentryALTinterwordspacing
K.~Miyamoto, H.~Tanaka, and S.~Nakamura, ``Meta-learning for emotion prediction from eeg while listening to music,'' in \emph{Companion Publication of the 2021 International Conference on Multimodal Interaction}, ser. ICMI '21 Companion.\hskip 1em plus 0.5em minus 0.4em\relax New York, NY, USA: Association for Computing Machinery, 2021, p. 324–328. [Online]. Available: \url{https://doi.org/10.1145/3461615.3486569}
\BIBentrySTDinterwordspacing

\bibitem{mamlbci2}
N.~Banluesombatkul, P.~Ouppaphan, P.~Leelaarporn, P.~Lakhan, B.~Chaitusaney, N.~Jaimchariyatam, E.~Chuangsuwanich, W.~Chen, H.~Phan, N.~Dilokthanakul, and T.~Wilaiprasitporn, ``Metasleeplearner: A pilot study on fast adaptation of bio-signals-based sleep stage classifier to new individual subject using meta-learning,'' \emph{IEEE Journal of Biomedical and Health Informatics}, vol.~25, no.~6, pp. 1949--1963, 2021.

\bibitem{mamlbci3}
\BIBentryALTinterwordspacing
L.~Chen, Z.~Yu, and J.~Yang, ``Spd-cnn: A plain cnn-based model using the symmetric positive definite matrices for cross-subject eeg classification with meta-transfer-learning,'' \emph{Frontiers in Neurorobotics}, vol.~16, 2022. [Online]. Available: \url{https://www.frontiersin.org/articles/10.3389/fnbot.2022.958052}
\BIBentrySTDinterwordspacing

\bibitem{mamlbci4}
\BIBentryALTinterwordspacing
J.~Li, F.~Wang, H.~Huang, F.~Qi, and J.~Pan, ``A novel semi-supervised meta learning method for subject-transfer brain–computer interface,'' \emph{Neural Networks}, vol. 163, pp. 195--204, 2023. [Online]. Available: \url{https://www.sciencedirect.com/science/article/pii/S0893608023001740}
\BIBentrySTDinterwordspacing

\bibitem{eegnet}
\BIBentryALTinterwordspacing
V.~J. Lawhern, A.~J. Solon, N.~R. Waytowich, S.~M. Gordon, C.~P. Hung, and B.~J. Lance, ``Eegnet: a compact convolutional neural network for eeg-based brain–computer interfaces,'' \emph{Journal of Neural Engineering}, vol.~15, no.~5, p. 056013, jul 2018. [Online]. Available: \url{https://dx.doi.org/10.1088/1741-2552/aace8c}
\BIBentrySTDinterwordspacing

\bibitem{transformer_1}
J.~Xie, J.~Zhang, J.~Sun, Z.~Ma, L.~Qin, G.~Li, H.~Zhou, and Y.~Zhan, ``A transformer-based approach combining deep learning network and spatial-temporal information for raw eeg classification,'' \emph{IEEE Transactions on Neural Systems and Rehabilitation Engineering}, vol.~30, pp. 2126--2136, 2022.

\bibitem{transformer_2}
\BIBentryALTinterwordspacing
D.~Kostas, S.~Aroca-Ouellette, and F.~Rudzicz, ``Bendr: Using transformers and a contrastive self-supervised learning task to learn from massive amounts of eeg data,'' \emph{Frontiers in Human Neuroscience}, vol.~15, 2021. [Online]. Available: \url{https://www.frontiersin.org/articles/10.3389/fnhum.2021.653659}
\BIBentrySTDinterwordspacing

\bibitem{transformer_3}
\BIBentryALTinterwordspacing
H.-Y.~S. Chien, H.~Goh, C.~M. Sandino, and J.~Y. Cheng, ``Maeeg: Masked auto-encoder for eeg representation learning,'' in \emph{NeurIPS Workshop}, 2022. [Online]. Available: \url{https://arxiv.org/abs/2211.02625}
\BIBentrySTDinterwordspacing

\bibitem{pytorch}
\BIBentryALTinterwordspacing
A.~Paszke, S.~Gross, F.~Massa, A.~Lerer, J.~Bradbury, G.~Chanan, T.~Killeen, Z.~Lin, N.~Gimelshein, L.~Antiga, A.~Desmaison, A.~Kopf, E.~Yang, Z.~DeVito, M.~Raison, A.~Tejani, S.~Chilamkurthy, B.~Steiner, L.~Fang, J.~Bai, and S.~Chintala, ``Pytorch: An imperative style, high-performance deep learning library,'' in \emph{Advances in Neural Information Processing Systems 32}.\hskip 1em plus 0.5em minus 0.4em\relax Curran Associates, Inc., 2019, pp. 8024--8035. [Online]. Available: \url{http://papers.neurips.cc/paper/9015-pytorch-an-imperative-style-high-performance-deep-learning-library.pdf}
\BIBentrySTDinterwordspacing

\bibitem{MAML_Pytorch}
L.~Long, ``Maml-pytorch implementation,'' \url{https://github.com/dragen1860/MAML-Pytorch}, 2018.

\bibitem{physionet}
A.~L. Goldberger, L.~A. Amaral, L.~Glass, J.~M. Hausdorff, P.~C. Ivanov, R.~G. Mark, J.~E. Mietus, G.~B. Moody, C.-K. Peng, and H.~E. Stanley, ``Physiobank, physiotoolkit, and physionet: components of a new research resource for complex physiologic signals,'' \emph{circulation}, vol. 101, no.~23, pp. e215--e220, 2000.

\bibitem{mne}
A.~Gramfort, M.~Luessi, E.~Larson, D.~A. Engemann, D.~Strohmeier, C.~Brodbeck, R.~Goj, M.~Jas, T.~Brooks, L.~Parkkonen, and M.~S. H{\"a}m{\"a}l{\"a}inen, ``{{MEG}} and {{EEG}} data analysis with {{MNE}}-{{Python}},'' \emph{Frontiers in Neuroscience}, vol.~7, no. 267, pp. 1--13, 2013.

\bibitem{gamma}
\BIBentryALTinterwordspacing
J.~L. Ulloa, ``The control of movements via motor gamma oscillations,'' \emph{Frontiers in Human Neuroscience}, vol.~15, 2022. [Online]. Available: \url{https://www.frontiersin.org/articles/10.3389/fnhum.2021.787157}
\BIBentrySTDinterwordspacing

\end{thebibliography}
\vspace{12pt}

\end{document}